\documentclass[pre,showpacs,twocolumn,superscriptaddress]{revtex4-1}
\usepackage{amssymb,graphicx,amsbsy,amsmath}

\begin{document}
\title{Zero-one-only process: a correlated random walk with a
stochastic ratchet}
\author{Seung Ki Baek}
\affiliation{Department of Physics, Pukyong National University, Busan 608-737,
Korea}
\author{Hawoong Jeong}
\affiliation{Department of Physics, Korea Advanced Institute of Science and
Technology, Daejeon 305-701, Korea}
\author{Seung-Woo Son}
\email[E-mail: ]{sonswoo@hanyang.ac.kr}
\affiliation{Department of Applied Physics, Hanyang University, Ansan 426-791,
Korea}
\author{Beom Jun Kim}
\email[E-mail: ]{beomjun@skku.edu}
\affiliation{BK21 Physics Research Division and Department of Physics,
Sungkyunkwan University, Suwon 440-746, Korea}

\begin{abstract}
The investigation of random walks is central to a variety of stochastic
processes in physics, chemistry, and biology. To describe a transport
phenomenon, we study a variant of the one-dimensional persistent random walk,
which we call a zero-one-only process. It makes a step in the same direction as
the previous step with probability $p$, and stops to change the direction with
$1-p$. By using the generating-function method, we calculate its characteristic
quantities such as the statistical moments and probability of the first return.
\end{abstract}

\pacs{05.40.Fb, 87.15.Vv, 02.10.Ox}
\maketitle

\section{Introduction}

The simple random walker is one of the most important conceptual tools
in statistical physics. It lies behind our understanding of diffusive motions
in thermal equilibrium, and almost every statistical estimate makes use of its
properties based on the central limit theorem~\cite{grinstead,rudnick}. It is
also useful in the biological context, because it explains many behavioral
aspects of micro-organisms swimming in a viscous liquid~\cite{purcell}.
Recent experiments report another biological application of
random walks, performed by repair proteins along one-dimensional DNA
sequences~\cite{dna1,dna2}.
The experimental results show the followings: First, the net displacements are
distributed symmetrically from the starting point. Second, the mean-squared
displacement is found to increase linearly with time. These are exactly the
characteristics of the simple random walk. Such a diffusive process is found
efficient both in energy and time: The proteins slide along DNA without
requiring ATP because the process is driven by thermal energy, and the group of
repair proteins in a cell would check the entire genome sequence within 3
minutes even if this one-dimensional diffusion was the only scanning
mechanism~\cite{dna1}.

There are also interesting variations of the simple random walk, one of which is
the persistent random walk~\cite{masoliver,boguna,weiss,rudnick}. The
persistent random walker has a `memory' or `momentum' in the sense that it takes
a step in the same direction as the previous one with probability, say,
$\alpha$, and in the opposite direction with $1-\alpha$. Such dynamics
introduces correlation in the walker's displacements. However, when we talk
about its `memory', it should be understood in a rather loose sense, because
this model can still be described by a second-order Markovian process, which is
essentially memoryless.

The precise way to introduce persistence depends on the detailed mechanism
that we are to describe. In this work, we consider a slightly different type of
a persistent random walker which has an external state as its position together
with an internal state that prescribes its direction. The walker receives as an
input a binary string composed of 1 and 0. The former bit acts on the
external state, moving the walker by one discrete step in the prescribed
direction. The latter, on the other hand, acts only on the internal state with
flipping the direction. The point is that there occurs no displacement
in the latter situation, differing from the conventional persistent random walk.
Therefore, in terms of the displacement, there are three possibilities, i.e.,
-1, 0, and +1, at each time step.
Such persistence as considered in this work due to the separation of internal
and external variables is actually possible in some transport phenomena, for
example, if the walker has a ratchet, which forces the motion in a particular
direction but is controllable by inputs from outside.
If the input string is random so that it contains 1
with probability $p$ and 0 with probability $q\equiv 1-p$, our model can be
analyzed by solving a second-order Markovian process with the
generation-function method. In addition, it is also possible to obtain the
generating function for the returning probability in a closed form at
$p=q=\frac{1}{2}$.

This work is organized as follows: In the next section, we present analytic
results for the movement of this random walker by using the
generating-function method.
How it returns to the starting point will be discussed in
Sec.~\ref{sec:return}.
After comparing the analytic
results with numerical ones, we conclude this work.

\section{Master equation}

Suppose that the walker wanders along the one-dimensional line from
$-\infty$ to $\infty$. Its position is represented by an integer $n$. We assume
that the walker starts from the origin, i.e., $n=0$, at time $t=0$. Every time
step, the walker reads a bit from an input string which we denote
as $\{X(t)\}$ with $X(t)=1$ at probability $p$ and $X(t)=0$ at probability
$q(=1-p)$ for $t\ge 1$.
The probability for the walker to occupy position $n$ at time $t$ is denoted as
$P^\pm(n,t)$, where the superscript means
the initial direction, $+$ or $-$, of the walker.
Our initial condition is such that the walker is located at the origin with the
positive direction, as expressed by $P^+(n,0) = \delta_{n0}$ and $P^-(n,0)=0$.
At time $t \ge 1$, we have
\begin{equation}
P^+(n,t) = p P^+(n-1,t-1) + q P^-(n,t-1),
\label{eq:p+}
\end{equation}
where the first term corresponds to the case with $X(1)=1$, which is
equivalent to shifting the walker by one lattice spacing.
The second term corresponds to the other case with
$X(1)=0$, which amounts to reverting the direction.
By the same logic, we have another recursion relation:
\begin{equation}
P^-(n,t) = p P^-(n+1,t-1) + q P^+(n,t-1).
\label{eq:p-}
\end{equation}
Let us define
\begin{equation}
Q^\pm(x,t) \equiv \sum_{n=-\infty}^{\infty} x^n P^\pm(n,t),
\label{eq:q+-}
\end{equation}
with the initial condition reexpressed as $Q^+(x,0) = 1$ and $Q^-(x,0)=0$.
In terms of Eq.~(\ref{eq:q+-}),
we write Eqs.~(\ref{eq:p+}) and (\ref{eq:p-}) as
\begin{equation}
\begin{bmatrix}
Q^+(x,t)\\ Q^-(x,t)
\end{bmatrix}
=
\begin{bmatrix}
px & q \\
q & px^{-1}
\end{bmatrix}
\begin{bmatrix}
Q^+(x,t-1)\\ Q^-(x,t-1)
\end{bmatrix}.
\label{eq:q2}
\end{equation}
The eigenvalues of the matrix are $\lambda_1 = \frac{p(x+x^{-1})-\sqrt{D}}{2}$
and $\lambda_2 = \frac{p(x+x^{-1})+\sqrt{D}}{2}$ with $D \equiv 4q^2 +
p^2(x-x^{-1})^2$.
A general expression for $Q^+(t)$ is obtained by diagonalizing
the $2 \times 2$ matrix in Eq.~(\ref{eq:q2}).
Considering the initial condition at $t=0$, we find that
\begin{eqnarray}
Q^+(x,t) &=& 2^{-1-t} \left\{ \left[1- \frac{2q+p(x-x^{-1})}{\sqrt{D}}\right]
\lambda_1^t \right.\nonumber\\
&&+ \left. \left[1+ \frac{2q+p(x-x^{-1})}{\sqrt{D}}\right] \lambda_2^t
\right\}.
\label{eq:q+t}
\end{eqnarray}
The mean and variance of the position are obtained as
\begin{equation}
\left<n \right> = \frac{p [1-(p-q)^t]}{2q}
\label{eq:mean}
\end{equation}
and
\begin{equation}
\sigma^2 = \frac{p \left\{ 4t-p\left[ 3-4(p-q)^t + (p-q)^{2t} + 4t \right]
\right\}}{4q^2}.
\label{eq:variance}
\end{equation}
In the limit of $t \rightarrow \infty$, we can approximate $\left< n \right>
\approx \frac{p}{2q}$ and $\sigma^2 \approx
\left(\frac{p}{q} \right) t$ for $0\le p <1$.
If $p=q=\frac{1}{2}$, in particular, the mean position is obtained as
$\left< n \right> = \frac{1}{2}$, and the variance is
$\sigma^2 = t-\frac{3}{4}$ at arbitrary $t$.
\begin{figure}
\includegraphics[width=0.45\textwidth]{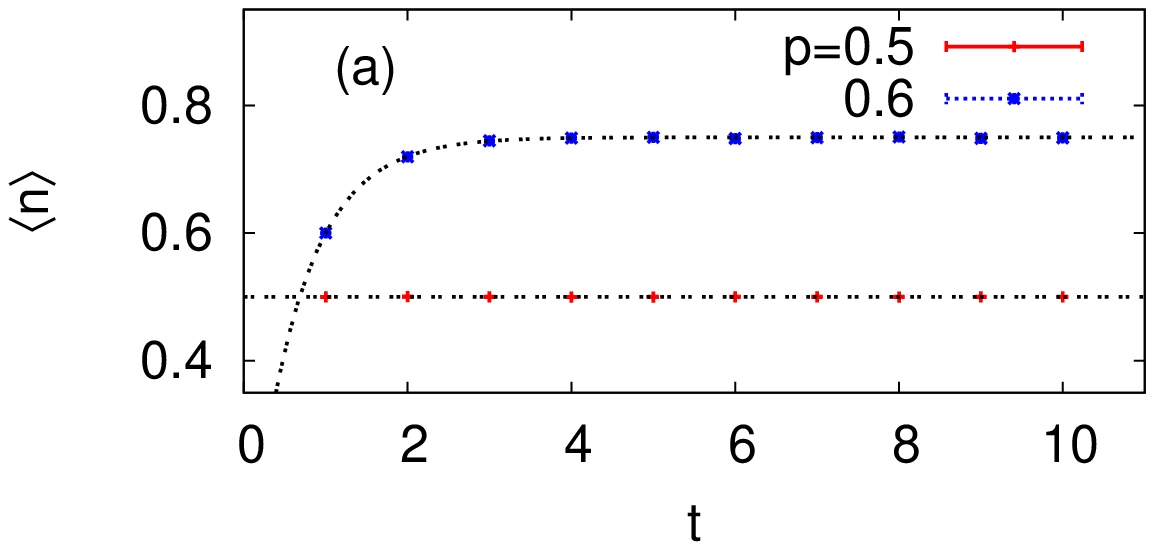}
\includegraphics[width=0.45\textwidth]{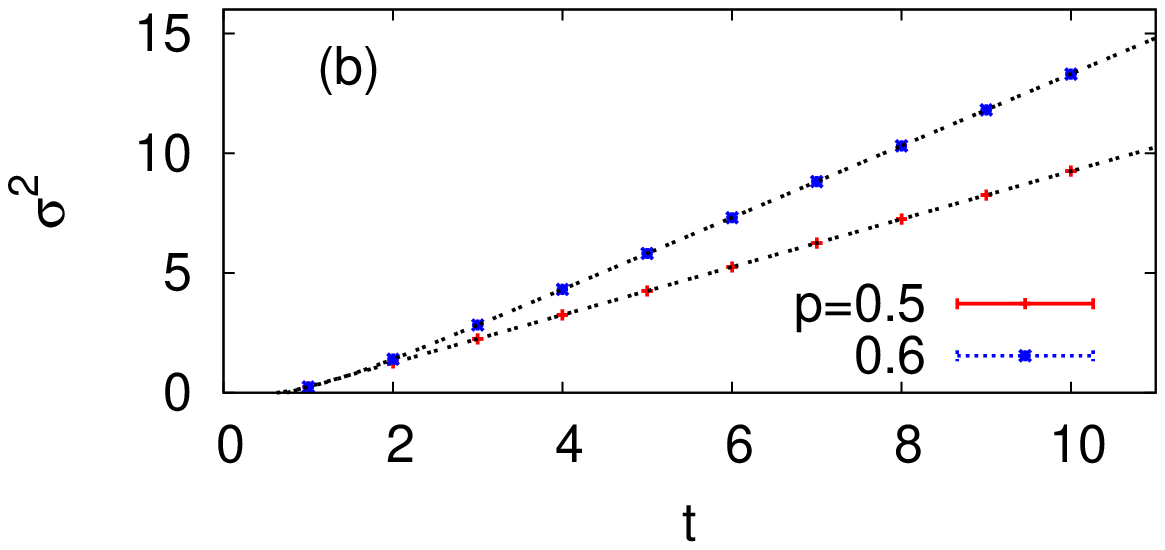}
\caption{(a) Mean and (b) variance of the position of our persistent random
walker at two different values of $p$ as time $t$ varies. The symbols are
averages over $3\times 10^7$ random samples, while the dotted lines
represent the analytic predictions in Eqs.~(\ref{eq:mean}) and
(\ref{eq:variance}).  Error bars are shown but not larger than the symbol
size.}
\label{fig:moments}
\end{figure}
Figure~\ref{fig:moments} depicts numerical results from our Monte Carlo
calculation over $3 \times 10^7$ samples,
compared with Eqs.~(\ref{eq:mean}) and (\ref{eq:variance}),
respectively.
The calculated mean and variance fully coincide with the analytic predictions.

\section{Return probability}
\label{sec:return}
Let us now consider the probability of return to the origin at time $t$ and
denote it as $u_t$. It is convenient to define $u_0 \equiv 1$.
If $t$ is odd, i.e., if $t = 2m+1$ with $m=0,1,\ldots$,
$u_t$ is then given as
\begin{eqnarray}
u_t 
&=& q \sum_{k=0}^m \binom{m}{k} \binom{m}{k} p^{2m-2k} q^{2k}\label{eq:utpq}\\
&=& q p^{2m} {}_2F_1 ( -m,-m;1;q^2/p^2),\nonumber
\end{eqnarray}
where $_2F_1$ is a hypergeometric function~\cite{hyper}.
The reason behind Eq.~(\ref{eq:utpq}) is roughly explained as
follows: The summand can be interpreted as pairing two strings of length $m$,
for each of which there are $k$ bits of $1$ and the other $(m-k)$ bits of $0$.
The probability $q$ in front of the right-hand side of Eq.~(\ref{eq:utpq})
means that for every such pair, one finds a proper place to insert $0$ so as to
bring the walker back to the origin.
If we define $b_t$ as the probability to occupy $n=-1$ at time $t$,
we can establish the following relation:
\begin{equation}
u_{t+1} = p b_t + q u_t,
\label{eq:ut+1}
\end{equation}
by conditioning the first bit, $X(1)$.
The probability $b_t$ turns out to be closely related to the Narayana
number~\cite{combi}:
\begin{equation}
N(m,k) = \frac{1}{m} \binom{m}{k} \binom{m}{k-1},
\end{equation}
which describes the number of possibilities to have $m$ pairs of correctly
matched parentheses with $k$ distinct nestings. If $t=2m+1$, in particular, we
find an explicit expression for $b_t$ as
\begin{eqnarray}
b_t 
&=& p \sum_{k=1}^m \binom{m}{k} \binom{m}{k-1}  p^{2m-2k} q^{2k}
\label{eq:bt}\\
&=& mp^{2m-1}q^2 {}_2F_1(1-m,-m;2;q^2/p^2).\nonumber
\end{eqnarray}
Plugging Eqs.~(\ref{eq:utpq}) and (\ref{eq:bt}) into Eq.~(\ref{eq:ut+1}), we
have an expression for $u_t$ when $t$ is even, too.

Equation~(\ref{eq:utpq}) simplifies for $p=q=\frac{1}{2}$ due to the
following identity in combinatorics:
\begin{equation}
\sum_{k=0}^m \binom{m}{k}^2 = \binom{2m}{m}.
\end{equation}
We restrict ourselves to this specific case of $p=q=\frac{1}{2}$ henceforth.
Then, the return probability $u_t$ found above obeys the following recursion
relation:
\begin{equation}
t u_t = u_{t-1} + (t-2) u_{t-2}.
\label{eq:return}
\end{equation}
Note that $u_1 = \frac{1}{2}$ because one should get $X(1)=0$ to stay at the
origin. Together with $u_0 \equiv 1$, one can find $u_t$ recursively by using
Eq.~(\ref{eq:return}). However, it is more useful to introduce
the generating function for $u_t$ as
\begin{equation}
U(x) \equiv \sum_{t=0}^\infty u_t x^t,
\end{equation}
and Eq.~(\ref{eq:return}) then reduces to
\begin{equation}
\frac{d}{dx}U(x) = U(x)-\frac{1}{2} + x^2 \frac{d}{dx}U(x).
\end{equation}
This ordinary differential equation is solved to yield
\begin{equation}
U(x) = \frac{1}{2} \left( \sqrt{\frac{1+x}{1-x}} +1 \right)
= 1 + \frac{x}{2} + \frac{x^2}{4} + \frac{x^3}{4} + \ldots,
\label{eq:ux}
\end{equation}
from which one can extract $u_t$ at arbitrary $t$.

\begin{figure}
\includegraphics[width=0.5\textwidth]{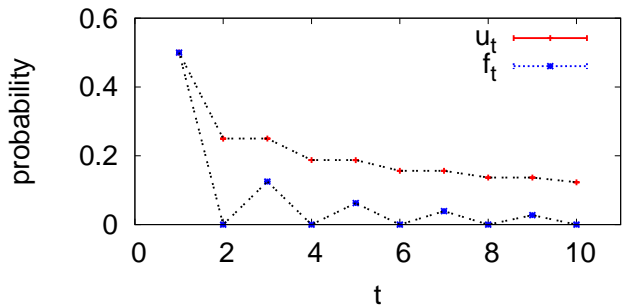}
\caption{Probability $u_t$ to return to the origin,
and probability $f_t$ for the first return at time $t$.
We have fixed the parameter
$p$ at $\frac{1}{2}$ to compare the numerical results over $3 \times 10^7$
random samples (symbols) with the analytic ones (dotted) from
Eqs.~(\ref{eq:ux}) and (\ref{eq:fx}).
Error bars are shown but smaller than the symbol size.}
\label{fig:return}
\end{figure}
We may furthermore define $f_t$ as the probability of the first return to the
origin at time $t$ with setting $f_0 \equiv 0$. It is a standard
exercise~\cite{grinstead} to decompose $u_{t>0}$ into
\begin{equation}
u_t = f_0 u_t + f_1 u_{t-1} + \ldots + f_t u_0 = \sum_{\tau=0}^t
u_{t-\tau}f_\tau,
\label{eq:dec}
\end{equation}
according to the first return time.
Denoting the generating function for $f_t$ as
\begin{equation}
F(x) \equiv \sum_{t=0}^\infty f_t x^t,
\end{equation}
we may rewrite Eq.~(\ref{eq:dec}) as
\begin{eqnarray}
U(x) &=& 1 + \sum_{t=0}^\infty \sum_{\tau=0}^t u_{t-\tau} f_\tau x^t\nonumber\\
&=& 1 + \sum_{\tau=0}^\infty f_\tau x^\tau \sum_{t=\tau}^\infty u_{t-\tau}
x^{t-\tau}\label{eq:conv}\\
&=& 1 + F(x) U(x).\nonumber
\end{eqnarray}
We thus obtain
\begin{eqnarray}
F(x) &=& 1 - \frac{1}{U(x)} = \frac{\sqrt{1+x} - \sqrt{1-x}}{\sqrt{1+x} +
\sqrt{1-x}} \label{eq:fx}\\
&=& \frac{x}{2} + \frac{x^3}{8} + \frac{x^5}{16} + \frac{5x^7}{128} +
\frac{7x^9}{256} + \ldots,\nonumber
\end{eqnarray}
from which one can read off $f_t$ at any $t$.
Note that $F(x)$ approaches one as $x \rightarrow 1$,
which means that this walk is recurrent.
Note also that Eq.~(\ref{eq:fx}) is an odd function, because the first return
after leaving the origin is impossible when $t$ is even.
We have checked the expressions for $u_t$ and $f_t$ with Monte Carlo
calculation and drawn the results in Fig.~\ref{fig:return}.

There is another useful expression for $f_t$ written as
\begin{equation}
f_t = \frac{[1-(-1)^t] \Gamma\left( \frac{t}{2} \right)}{4\sqrt{\pi}
~\Gamma\left( \frac{t}{2}+\frac{3}{2} \right)},
\label{eq:ft}
\end{equation}
whereby we can obtain an important quantity $g_t$, defined as the number of
returns to the origin by time $t$. We decompose $g_t$ according to the
first return time $\tau$: If the first return time is $\tau$, it means that
there must be at least $2^t f_\tau$ returns among the $2^t$ possible
trajectories in total.  In addition, when those trajectories hit the origin at
$\tau$, the number of possibilities should be $2^\tau f_\tau$, each of which may
contain more returning events during the remaining $t-\tau$ time steps. This can
be expressed as $2^\tau f_\tau g_{t-\tau}$ by the definition of $g_t$. To sum
up, $f_t$ and $g_t$ are related by
\begin{equation}
g_t = \sum_{\tau=1}^t \left( f_\tau 2^t + 2^\tau f_\tau g_{t-\tau} \right).
\label{eq:gt}
\end{equation}
Let us consider the corresponding generating function:
\begin{eqnarray}
G(x) &\equiv& \sum_{t=0}^\infty g_t x^t
= \sum_{t=0}^\infty \sum_{\tau=1}^t \left( f_\tau 2^t + 2^\tau f_\tau
g_{t-\tau} \right) x^t\\
&=& \sum_{t=0}^\infty (2x)^t \sum_{\tau=1}^t f_\tau + \sum_{t=0}^\infty
\sum_{\tau=1}^t  f_\tau (2x)^\tau \times g_{t-\tau} x^{t-\tau},\nonumber
\end{eqnarray}
where the first term can be evaluated by using the general expression in
Eq.~(\ref{eq:ft}) and the second term can be expressed by convolution as in
Eq.~(\ref{eq:conv}). After some algebra, we arrive at
\begin{equation}
G(x) = \frac{1-\sqrt{1-4x^2}}{2x(1-2x)} + F(2x) G(x),
\end{equation}
which yields
\begin{eqnarray}
G(x) &=& \frac{1-\sqrt{1-4x^2}}{2x(1-2x) [1-F(2x)]}\\
&=& x + 3x^2 + 8x^3 + 19x^4 + 44x^5 + \ldots.\nonumber
\end{eqnarray}
For example, we find three returning events by time $t=2$, because a trajectory
generated by $00$ visits the origin twice and another trajectory from $01$
does it once, while the other two are kept away from the origin.
The singular point of $G(x)$ at $x=\frac{1}{2}$ is of particular interest,
because $g_t \left( \frac{1}{2} \right)^t$ corresponds to the average number of
returns among all paths by time $t$. Rewriting
\begin{eqnarray}
G(x) = 2 U'(2x) - \frac{1}{2} \left( \frac{1}{1-2x} +
\frac{1}{\sqrt{1-4x^2}} \right),
\end{eqnarray}
where the prime denotes differentiation, we find that the singularity is
dominated by the first term on the right-hand side. As a result, if $t=2m+1 \gg
1$, the average number of returns scales as
\begin{equation}
\frac{g_t}{2^t} \sim \frac{m}{2^{2m}} \binom{2m}{m} \sim \sqrt{m},
\end{equation}
which is similar to the case of the simple random walk~\cite{grinstead}.

\section{Conclusion}

In summary, we have considered a variant of the persistent random walk and
calculated its statistical properties by using the generating-function
method.
In the limit of $t \rightarrow \infty$, the distribution of the position
approaches the Gaussian function, and both the mean and variance are scaled by
the factor of $\frac{p}{q}$.
We have also obtained the generating functions for return events at
$p=q=\frac{1}{2}$ in closed forms.
The resulting analytic predictions are fully confirmed by numerical
simulation.

Before concluding this work, let us briefly mention how to deal with the
periodic boundary condition.
We may assume that there are $L$ sites, i.e., $n=1,2,\ldots,L$, with the
periodic boundary condition such that $P^\pm(n+L,t) = P^\pm(n,t)$. The
generating function method is made compatible with the boundary condition
if we introduce $z_k = e^{2\pi i k/L}$ with $k=1,2,\ldots,L$.
Then, the following expression
\begin{equation}
\tilde{Q}^+(k,t) \equiv \sum_{n=1}^L z_k^n P^+(n,t)
\end{equation}
corresponds to the discrete Fourier transform of $P^+(n,t)$. The inverse
transform therefore gives us
\begin{equation}
P^+(n,t) = \frac{1}{L} \sum_{k=1}^L z_k^{-n} \tilde{Q}^+(k,t).
\label{eq:ift}
\end{equation}
We have already obtained the expression $Q^+(z,t)$ in Eq.~(\ref{eq:q+t}) so we
need only to use $z_k$ instead of $z$ and then plug the resulting
$\tilde{Q}^+(k,t)$
into Eq.~(\ref{eq:ift}). As $t \rightarrow \infty$, only the lowest mode with
$k=L$ survives, which corresponds to $z_k = 1$. Since the conservation of total
probability automatically implies $Q^+(z=1,t) = 1$, we immediately see from
Eq.~(\ref{eq:ift}) that $P^+(n,t\rightarrow \infty) = 1/L$, that is, a uniform
distribution as indicated by our intuition.

\acknowledgments
This work was supported by a research grant of Pukyong National University
(2013).

%
\end{document}